\documentclass[aps,twocolumn,showpacs,footinbib]{revtex4}%
\usepackage[latin9]{inputenc}
\usepackage{amssymb}
\usepackage{amsmath}
\usepackage{amscd}
\usepackage{graphicx}
\usepackage{subfigure}
\usepackage{float}

\begin{document}
\draft
\title{Generation of entangled states for many multi-level atoms in a thermal
cavity and ions in thermal motion}
\author{Shi-Biao Zheng\thanks{%
E-mail: sbzheng@pub5.fz.fj.cn}}
\address{Department of Electronic Science and Applied Physics\\
Fuzhou University\\
Fuzhou 350002, P. R. China}
\date{\today }

\begin{abstract}
We propose a scheme for generating entangled states for two or more
multi-level atoms in a thermal cavity. The photon-number dependent parts in
the effective Hamiltonian are canceled with the assistant of a strong
classical field. Thus the scheme is insensitive to both the cavity decay and
the thermal field. The scheme does not require individual addressing of the
atoms in the cavity. The scheme can also be used to generate entangled
states for many hot multi-level ions.
\end{abstract}

 \maketitle \vskip
0.5cm

\narrowtext

Entanglement of two or more particles is not only of significance for test
of quantum mechanics against local hidden theory [1-3], but also useful in
quantum cryptography [4] and quantum teleportation [5]. Most of research in
quantum nonlocality and quantum information is based on entanglement of
two-level particles. Entangled states for two-level particles have been
observed for photons [6-8], atoms in cavity QED [9-11], and ions in a trap
[12-14].

Recently, it has been shown that violations of local realism by two
entangled N-dimensional systems are stronger than for two qubits [15]. The
Greenberger-Horne-Zeiliner paradox has also been extended to the many
N-dimensional systems [16]. Furthermore, it has been shown that quantum
cryptography based on entangled qutrits is more secure than that based on
entangled qubits [17]. High-dimensional entanglement for photons has been
observed [18-21]. However, there have no reports on the realization of
entanglement for multi-level massive particles. Recently, Zou et al. [22]
have proposed a scheme for the generation of entangled states for two
three-level atoms in cavity QED using nonresonant interaction of two atoms
with a cavity [23]. The scheme is insensitive to cavity decay. The main
drawback of the scheme is that it requires individual addressing of the
atoms when both atoms are still in the cavity, which is experimentally
problematic.

In this paper we propose a scheme for generating entangled states for many
multi-level atoms in cavity QED and ions in a trap. In cavity QED, our
scheme does not require individual addressing of the atoms in the cavity.
Another distinct feature of the present scheme is that the photon-number
dependent parts in the effective Hamiltonian are canceled with the assistant
of a strong classical driving field. Due to this feature the scheme is
insensitive to both the cavity decay and thermal field. For the trapped
ions, our scheme is insensitive to the thermal motion.

We consider N identical ladder-type three-level atoms simultaneously
interacting with a single-mode cavity field and driven by a classical field.
The atomic states are denoted by $\left| g\right\rangle $, $\left|
e\right\rangle ,$ and $\left| i\right\rangle .$ The transition frequency
between the states $\left| e\right\rangle $ and $\left| i\right\rangle $ is
highly detuned from the cavity frequency and thus the state $\left|
i\right\rangle $ is not affected during the atom-cavity interaction. The
Hamiltonian is (assuming $\hbar =1$) [24,25]
\begin{eqnarray}
&H=\omega _0\sum_{j=1}^NS_{z,j}+\omega
_aa^{+}a+\sum_{j=1}^N[g(a^{+}S_j^{-}+aS_j^{+})&\cr\cr&+\Omega
(S_j^{+}e^{-i\omega t}+S_j^{-}e^{i\omega t})]&,
\end{eqnarray}
where S$_j^{+}=\left| e_j\right\rangle \left\langle g_j\right| $, S$%
_j^{-}=\left| g_j\right\rangle \left\langle e_j\right| ,$ $S_{z,j}=\frac 12%
(\left| e_j\right\rangle \left\langle e_j\right| -\left| g_j\right\rangle
\left\langle g_j\right| ),$ with $\left| e_j\right\rangle $ and $\left|
g_j\right\rangle $ (j=1,2) being the excited and ground states of the jth
atom, a$^{+}$ and a are the creation and annihilation operators for the
cavity mode, and g is the atom-cavity coupling strength, and $\Omega $ is
the Rabi frequency of the classical field. Assume that $\omega _0=\omega .$
Then the interaction Hamiltonian, in the interaction picture, is
\begin{equation}
H_i=\sum_{j=1}^N[g(e^{-i\delta t}a^{+}S_j^{-}+e^{i\delta t}aS_j^{+})+\Omega
(S_j^{+}+S_j^{-})],
\end{equation}
$\delta $ is the detuning between the atomic transition frequency $\omega _0$
and cavity frequency $\omega .$ Define the new atomic basis

\begin{equation}
\left| +_j\right\rangle =\frac 1{\sqrt{2}}(\left| g_j\right\rangle +\left|
e_j\right\rangle ),\text{ }\left| -_j\right\rangle =\frac 1{\sqrt{2}}(\left|
g_j\right\rangle -\left| e_j\right\rangle ).
\end{equation}
Then we can rewrite $H_i$ as
\begin{eqnarray}
H_i=\sum_{j=1}^N[ge^{-i\delta t}a^{+}(\sigma _{z,j}+\frac 12\sigma _j^{+}-%
\frac 12\sigma _j^{-})\cr\cr+e^{i\delta t}a(\sigma _{z,j}+\frac 12\sigma _j^{-}-%
\frac 12\sigma _j^{+})+2\Omega \sigma _{z,j}],
\end{eqnarray}
where $\sigma _{z,j}=\frac 12(\left| +_j\right\rangle \left\langle
+_j\right| -\left| -_j\right\rangle \left\langle -_j\right| ),$
$\sigma _j^{+}=\left| +_j\right\rangle \left\langle -_j\right| $ and
$\sigma _j^{-}=\left| -_j\right\rangle \left\langle +_j\right| .$

The time evolution of this system is decided by Schr\"odinger's equation:
\begin{equation}
i\frac{d|\psi (t)\rangle }{dt}=H_i|\psi (t)\rangle .
\end{equation}
Perform the unitary transformation
\begin{equation}
|\psi (t)\rangle =e^{-iH_0t}|\psi ^{^{\prime }}(t)\rangle ,
\end{equation}
with
\begin{equation}
H_0=2\Omega \sum_{j=1}^N\sigma _{z,j}.
\end{equation}
Then we obtain
\begin{equation}
i\frac{d|\psi ^{^{\prime }}(t)\rangle }{dt}=H_i^{^{\prime }}|\psi ^{^{\prime
}}(t)\rangle ,
\end{equation}
where
\begin{eqnarray}
H_i^{^{\prime }}=\sum_{j=1,2}[ge^{-i\delta t}a^{+}(\sigma _{z,j}+\frac 12%
\sigma _j^{+}e^{2i\Omega t}-\frac 12\sigma _j^{-}e^{-2i\Omega
t})\cr\cr+e^{i\delta t}a(\sigma _{z,j}+\frac 12\sigma
_j^{-}e^{-2i\Omega t}-\frac 12\sigma _j^{+}e^{2i\Omega t}).
\end{eqnarray}

Assuming that $2\Omega \gg \delta ,g,$ we can neglect the terms oscillating
fast. Then $H_i^{^{\prime }}$ reduces to
\begin{eqnarray}
H_i^{^{\prime }} &=&\sum_{j=1}^Ng(e^{-i\delta t}a^{+}+e^{i\delta t}a)\sigma
_{z,j} \\
\ &=&\frac 12\sum_{j=1}^Ng(e^{-i\delta t}a^{+}+e^{i\delta
t}a)(S_j^{+}+S_j^{-}).  \nonumber
\end{eqnarray}
In the case $\delta \gg g/2$, there is no energy exchange between the atomic
system and the cavity. The resonant transitions are $\left|
e_jg_kn\right\rangle \longleftrightarrow \left| g_je_kn\right\rangle $ and $%
\left| e_je_kn\right\rangle \longleftrightarrow \left| g_jg_kn\right\rangle
. $ The transition $\left| e_jg_kn\right\rangle \longleftrightarrow \left|
g_je_kn\right\rangle $ is mediated by $\left| g_jg_kn\pm 1\right\rangle $
and $\left| e_je_kn\pm 1\right\rangle $. The contributions of $\left|
g_jg_kn\pm 1\right\rangle $ are equal to those of $\left| e_je_kn\pm
1\right\rangle $. The corresponding Rabi frequency is given by
\begin{eqnarray}
2\frac{\left\langle e_jg_kn\right| H_i^{^{\prime }}\left|
g_jg_kn+1\right\rangle \left\langle g_jg_kn+1\right| H_i^{^{\prime
}}\left| g_je_kn\right\rangle }\delta\cr\cr +2\frac{\left\langle
e_jg_kn\right| H_i^{^{\prime }}\left| g_jg_kn-1\right\rangle
\left\langle g_jg_kn-1\right| H_i^{^{\prime }}\left|
g_je_kn\right\rangle }{-\delta } =\frac{g^2}{2\delta }.\cr
\end{eqnarray}
Since the transition paths interfere destructively the Rabi
frequency is independent of the photon-number of the cavity mode.
The destructive interference of transition amplitudes was first
proposed for trapped ions [26, 27]. The Rabi frequency for $\left|
e_je_kn\right\rangle \longleftrightarrow \left| g_jg_kn\right\rangle
,$ mediated by $\left| e_jg_kn\pm 1\right\rangle $ and $\left|
g_je_kn\pm 1\right\rangle ,$ is also equal to $g^2/(2\delta ).$ The
Stark shift for the state $\left| e_j\right\rangle $ is
\begin{eqnarray}
\frac{\left\langle e_jn\right| H_i^{^{\prime }}\left|
g_jn+1\right\rangle \left\langle g_jn+1\right| H_i^{^{\prime
}}\left| e_jn\right\rangle }\delta \cr\cr +\frac{\left\langle
e_jn\right| H_i^{^{\prime }}\left| g_jn-1\right\rangle \left\langle
g_jn-1\right| H_i^{^{\prime }}\left| e_jn\right\rangle }{-\delta }
=\frac{g^2}{4\delta }. \cr
\end{eqnarray}
The Stark shift for $\left| g_j\right\rangle $ is also $g^2/(4\delta ).$ The
strong classical field induces the terms $g(e^{-i\delta
t}a^{+}S_j^{+}+e^{i\delta t}aS_j^{-})$, which result in the photon-number
dependent Stark shifts negative to those induced by $g(e^{-i\delta
t}a^{+}S_j^{-}+e^{i\delta t}aS_j^{+}).$ Thus the photon-number dependent
Stark shifts are also cancelled. Then the effective Hamiltonian is given by
\begin{eqnarray}
H_e=\lambda [\frac 12\sum_{j=1}^N(\left| e_j\right\rangle
\left\langle e_j\right| +\left| g_j\right\rangle \left\langle
g_j\right|
)\cr+\sum_{j,k=1}^N(S_j^{+}S_k^{+}+S_j^{+}S_k^{-}+H.c.)],j\neq k
\end{eqnarray}
where $\lambda =\frac{g^2}{2\delta }.$ The distinct feature of the
effective Hamiltonian is that it is independent of the photon-number
of the cavity field. Without the strong classical field, the Stark
shift terms are proportional to the photon number, and the terms
$S_j^{+}S_k^{+}+H.c$ do not exist. The evolution operator of the
system is given by

\begin{equation}
U(t)=e^{-iH_0t}e^{-iH_et}.
\end{equation}
We note the atomic state evolution operator $U(t)$ is independent of the
cavity field state, allowing it to be in a thermal state.

We first consider the case that N=2. Assume the two atoms are initially in
the state $\left| g_1\right\rangle \left| g_2\right\rangle $. After an
interaction time t$_1$ the state of the system is
\begin{eqnarray}
\left| g_1\right\rangle \left| g_2\right\rangle \longrightarrow
e^{-i\lambda t_1}\{\cos (\lambda t_1)[\cos \Omega t_1\left|
g_1\right\rangle -i\sin \Omega t_1\left| e_1\right\rangle ]\cr[\cos
\Omega t_1\left| g_2\right\rangle -i\sin \Omega t_1\left|
e_2\right\rangle ] \cr-i\sin (\lambda t_1)[\cos \Omega t_1\left|
e_1\right\rangle -i\sin \Omega t_1\left| g_1\right\rangle ]\cr[\cos
\Omega t_1\left| e_2\right\rangle -i\sin \Omega t_1\left|
g_2\right\rangle ]\}.\cr
\end{eqnarray}
Choose the interaction time t$_1$ and Rabi frequency $\Omega $ appropriately
so that $\sin (\lambda t_1)=1/\sqrt{3}$ and $\Omega t_1=k\pi ,$with k being
an integer$.$ Then we have
\begin{equation}
\left| g_1\right\rangle \left| g_2\right\rangle \longrightarrow e^{-i\lambda
t_1}\{\sqrt{\frac 23}\left| g_1\right\rangle \left| g_2\right\rangle -i\frac
1{\sqrt{3}}\left| e_1\right\rangle \left| e_2\right\rangle \}.
\end{equation}

Now we switch off the classical field tuned to the $\left| g\right\rangle
\rightarrow \left| e\right\rangle ,$ and switch on another classical field
tuned to the $\left| e\right\rangle \rightarrow \left| f\right\rangle .$
Choosing the Rabi frequency and interaction time appropriately so that the
atoms undergoes the transitions: $\left| e\right\rangle \rightarrow \left|
f\right\rangle $. We here assume that this classical field is sufficiently
strong and thus the interaction time is so short that the dispersive
atom-cavity interaction can be neglected during the application of this
classical field. This leads to
\begin{equation}
e^{-i\lambda t_1}\{\sqrt{\frac 23}\left| g_1\right\rangle \left|
g_2\right\rangle -i\frac 1{\sqrt{3}}\left| f_1\right\rangle \left|
f_2\right\rangle \}.
\end{equation}

Then we again switch on the classical field tuned to the $\left|
g\right\rangle \rightarrow \left| e\right\rangle ,$ and switch off the field
tuned to the $\left| e\right\rangle \rightarrow \left| f\right\rangle .$ The
Hamiltonian is again given by Eq.(13). After another interaction time t$_2.$
We obtain
\begin{eqnarray}
\ e^{-i\lambda (t_1+t_2)}\sqrt{\frac 23}\{\cos (\lambda t_2)[\cos
\Omega ^{^{\prime }}t_2\left| g_1\right\rangle -i\sin \Omega
^{^{\prime }}t_2\left| e_1\right\rangle ]\cr[\cos \Omega ^{^{\prime
}}t_2\left| g_2\right\rangle -i\sin \Omega ^{^{\prime }}t_2\left|
e_2\right\rangle ] \cr -i\sin (\lambda t_2)[\cos \Omega ^{^{\prime
}}t_2\left| e_1\right\rangle -i\sin \Omega ^{^{\prime }}t_2\left|
g_1\right\rangle ]\cr[\cos \Omega ^{^{\prime }}t_2\left|
e_2\right\rangle -i\sin \Omega ^{^{\prime }}t_2\left|
g_2\right\rangle ]\} \cr\ -ie^{-i\lambda t_1}\frac 1{\sqrt{3}}\left|
f_1\right\rangle \left| f_2\right\rangle , \cr
\end{eqnarray}
where $\Omega ^{^{\prime }}$ is the Rabi frequency of the classical field
during the interaction time t$_2.$ Choose the interaction time t$_2$ and
Rabi frequency $\Omega ^{^{\prime }}$ appropriately so that $\lambda t_2=\pi
/4$ and $\Omega ^{^{\prime }}t_2=2k^{^{\prime }}\pi ,$with k$^{^{\prime }}$
being an integer$.$ Then we have
\begin{equation}
e^{-i\lambda t_1}\sqrt{\frac 13}\{e^{-i\lambda t_2}\left| g_1\right\rangle
\left| g_2\right\rangle -ie^{-i\lambda t_2}\left| e_1\right\rangle \left|
e_2\right\rangle -i\left| f_1\right\rangle \left| f_2\right\rangle \}.
\end{equation}
This is a maximally entangled state for the two three-level atoms. We here
do not require individually addressing of the atoms when they are in the
cavity. Furthermore, our scheme is not only insensitive to the cavity decay
but also insensitive to the thermal photons. The thermal field gradually
builds up during the operations [10]. Thus, our scheme is important in view
of experiment.

We now turn to the problem of generating entanglement for three or more
three-level atoms with a thermal cavity. The effective Hamiltonian H$_e$ can
also be rewritten as

\begin{equation}
H_e=2\lambda S_x^2,
\end{equation}
where

\begin{equation}
S_x=\frac 12\sum_{j=1}^N(S_j^{+}+S_j^{-}).
\end{equation}
Assume that the atoms are initially in the state $\left|
g_1g_2...g_N\right\rangle .$ Using the representation of the operator S$_z$,
the atomic state $\left| g_1g_2...g_N\right\rangle $ and $\left|
e_1e_2...e_N\right\rangle $ can be expressed as $\left|
N/2,-N/2\right\rangle $ and $\left| N/2,N/2\right\rangle .$ On the other
hand, such states can be expanded in terms of the eigenstates of S$_x$
[25,26,28]

\begin{equation}
\left| N/2,-N/2\right\rangle =\sum_{M=-N/2}^{N/2}C_M\left|
N/2,M\right\rangle _x.
\end{equation}
\begin{equation}
\left| N/2,N/2\right\rangle =\sum_{M=-N/2}^{N/2}C_M(-1)^{N/2-M}\left|
N/2,M\right\rangle _x.
\end{equation}
Thus, the evolution of the system is

\begin{equation}
\sum_{M=-N/2}^{N/2}C_Me^{-2i(\Omega M+\lambda M^2)t}\left|
N/2,M\right\rangle _x.
\end{equation}
When N is even M is an integer. With the choice $\lambda t=\pi /4$ and $%
\Omega t=n\pi $ we obtain
\begin{equation}
\begin{array}{c}
\frac 1{\sqrt{2}}\sum_{M=-N/2}^{N/2}C_M[e^{-i\pi /4}+e^{i\pi
/4}(-1)^M]\left| N/2,M\right\rangle _x \\
=\frac 1{\sqrt{2}}(e^{-i\pi /4}\left| g_1g_2...g_N\right\rangle +e^{i\pi
/4}(-1)^{N/2}\left| e_1e_2...e_N\right\rangle ).
\end{array}
\end{equation}
On the other hand, for the case that N is odd we choose $\lambda t=\pi /4$
and $\Omega t=(2n+\frac 34)\pi $. Then we obtain
\begin{equation}
\frac 1{\sqrt{2}}e^{i\frac 78\pi }[e^{-i\pi /4}\left|
g_1g_2...g_N\right\rangle +e^{i\pi /4}(-1)^{(1+N)/2}\left|
g_1g_2...g_N\right\rangle ].
\end{equation}
By this way we obtain a multiatom Greenberger-Horne-Zeilinger state [2].

We here assume that N is even. After the state of Eq. (26) is prepared we
switch off the classical field tuned to $\left| e\right\rangle \rightarrow
\left| g\right\rangle $ and perform the transformation: $\left|
e\right\rangle \rightarrow \left| f\right\rangle $. Then we have
\begin{equation}
\begin{array}{c}
\frac 1{\sqrt{2}}(e^{-i\pi /4}\left| g_1g_2...g_N\right\rangle +e^{i\pi
/4}(-1)^{N/2}\left| f_1f_2...f_N\right\rangle ).
\end{array}
\end{equation}
Then we again switch on the classical field tuned to $\left| e\right\rangle
\rightarrow \left| g\right\rangle $. After another interaction time t we
obtain an entangled state for the N three-level atoms
\begin{equation}
\begin{array}{c}
\frac 12e^{-i\pi /4}(e^{-i\pi /4}\left| g_1g_2...g_N\right\rangle
+e^{i\pi /4}(-1)^{N/2}\left| e_1e_2...e_N\right\rangle )\cr\cr+\frac
1{\sqrt{2}}e^{i\pi /4}(-1)^{N/2}\left| f_1f_2...f_N\right\rangle .
\end{array}
\end{equation}
After the N atoms exits the cavity we can prepare N-1 atoms into a maximally
entangled state via manipulating the Nth atom. We first perform the
transformations:

\begin{eqnarray}
\left| g_N\right\rangle &\rightarrow &\frac 1{\sqrt{2}}\left|
g_N\right\rangle +\frac 1{\sqrt{10}}\left| e_N\right\rangle -\sqrt{\frac 25}%
\left| f_N\right\rangle , \cr \left| e_N\right\rangle &\rightarrow
&-\frac 1{\sqrt{2}}\left|
g_N\right\rangle +\frac 1{\sqrt{10}}\left| e_N\right\rangle -\sqrt{\frac 25}%
\left| f_N\right\rangle ,  \cr \left| f_N\right\rangle &\rightarrow
&\frac 2{\sqrt{5}}\left| e_N\right\rangle +\sqrt{\frac 15}\left|
f_N\right\rangle .
\end{eqnarray}
Then we detect the state of the Nth atom. The detection of the state $\left|
f_N\right\rangle $ collapses the N-1 atoms onto the maximally entangled
state
\begin{equation}
\begin{array}{c}
\frac 1{\sqrt{3}}(e^{-i\pi /2}\left| g_1g_2...g_{N-1}\right\rangle
+(-1)^{N/2}\left| e_1e_2...e_{N-1}\right\rangle \cr\cr-e^{i\pi
/4}(-1)^{N/2}\left| f_1f_2...f_{N-1}\right\rangle .
\end{array}
\end{equation}
The probability of success is 0.3.

We note we can generate a maximally entangled state for N four-level atoms
determinately. The fourth level is $\left| h\right\rangle $. After the atoms
are prepared in the state of Eq. (28) we perform the transformation: $\left|
g\right\rangle \longleftrightarrow \left| f\right\rangle $ and $\left|
e\right\rangle \longleftrightarrow \left| h\right\rangle .$ This leads to
\begin{equation}
\begin{array}{c}
\frac 1{\sqrt{2}}e^{i\pi /4}(-1)^{N/2}\left| g_1g_2...g_N\right\rangle +%
\frac 12(e^{-i\pi /2}\left| f_1f_2...f_N\right\rangle\cr\cr
+(-1)^{N/2}\left| h_1h_2...h_N\right\rangle ).
\end{array}
\end{equation}
Then we again switch on the classical field tuned to $\left| e\right\rangle
\rightarrow \left| g\right\rangle $. After another interaction time t we
obtain an entangled state for the N four-level atoms
\begin{equation}
\begin{array}{c}
\frac 12[(-1)^{N/2}\left| g_1g_2...g_N\right\rangle +e^{i\pi
/2}\left| e_1e_2...e_N\right\rangle \cr +e^{-i\pi /2}\left|
f_1f_2...f_N\right\rangle +(-1)^{N/2}\left|
h_1h_2...h_N\right\rangle ].
\end{array}
\end{equation}

We note the idea can also be used to the ion trap system. We consider that N
ions are confined in a linear trap. Then we simultaneously excite the ions
with two lasers of frequencies $\omega _0+\nu +\delta $ and $\omega _0-\nu
-\delta $, where $\omega _0$ is the frequency of the transition $\left|
e\right\rangle \rightarrow \left| g\right\rangle $ and $\nu $ is the
frequency of the one collective vibrational mode. Suppose $\delta $ is much
smaller than $\nu $ and thus we can neglect other vibrational modes. In this
case the Hamiltonian for the system is given by [26,27]

\begin{equation}
\begin{array}{c}
\stackrel{\wedge }{H}=\nu \stackrel{\wedge }{a}^{+}\stackrel{\wedge }{a}%
+\omega _0\sum_{j=1}^N\stackrel{\wedge }{S}_{z,j}
+\{\Omega e^{-i\phi }\sum_{j=1}^N\stackrel{\wedge }{S}_j^{+}e^{i\eta (%
\stackrel{\wedge }{a}^{+}+\stackrel{\wedge }{a})}\cr\cr[e^{-i(\omega
_0+\nu +\delta )t}+e^{-i(\omega _0-\nu -\delta )t}]+H.c.\},
\end{array}
\label{7}
\end{equation}
where $\stackrel{\wedge }{a}^{+}$ and $\stackrel{\wedge }{a}$ are the
creation and annihilation operators for the collective vibrational mode, and
$\eta =k/\sqrt{2\nu M}$ is the Lamb-Dicke parameter with k being the
wavevector along the trap axis and M the mass of the ion collection. We here
have assumed the lasers have the same Rabi frequency $\Omega $, phase $\phi $
and same wavevector k. Furthermore, we consider the resolved sideband
regime, where the vibrational frequency $\nu $ is much larger than other
characteristic frequencies of the problem. In this case we discard the
rapidly oscillating terms and obtain the Hamiltonian in the interaction
picture
\begin{equation}
\begin{array}{c}
\stackrel{\wedge }{H}=\Omega e^{-\eta ^2/2}e^{-i\phi }\sum_{j=1}^N\stackrel{%
\wedge }{S}_j^{+}\sum_{j=0}^\infty \frac{(i\eta
)^{2j+1}}{j!(j+1)!}\cr\cr\left[
\stackrel{\wedge }{a}^{+(j+1)}\stackrel{\wedge }{a}^je^{-i\delta t}+%
\stackrel{\wedge }{a}^{+j}\stackrel{\wedge }{a}^{j+1}e^{i\delta
t}\right] +H.c.,
\end{array}
\label{8}
\end{equation}
In the Lamb-Dicke regime, i.e., $\eta \sqrt{n+1}\ll 1$ with n being the
phonon number, the Hamiltonian of Eq.(34) can be approximated by the
expansion to the first order in $\eta $%
\begin{equation}
\stackrel{\wedge }{H}=i\eta \Omega e^{-i\phi }\sum_{j=1}^N\stackrel{\wedge }{%
S}_j^{+}(\stackrel{\wedge }{a}^{+}e^{-i\delta t}+\stackrel{\wedge }{a}%
e^{i\delta t})+H.c.,  \label{8}
\end{equation}
When $\delta \gg \eta \Omega $ and $\phi =\pi /2$ the effective Hamiltonian
has the same form as Eq. (13), with $\lambda =2\Omega ^2\eta ^2/\delta .$
For the case N=2 we focus the two lasers on the ions for a time arc$\sin (1/%
\sqrt{3})/\lambda ,$ then perform the transformation: $\left| e\right\rangle
\rightarrow \left| f\right\rangle $ with $\left| f\right\rangle $ being
another internal state, followed by the application of the above mentioned
two lasers for a time $\pi /(4\lambda )$. The two ions are prepared in the
state of Eq. (19). Using the procedure similar to that for cavity QED we can
also generate entangled states for many multi-level ions. The effective
Hamiltonian does not involve the external degree of freedom and thus the
scheme is insensitive to the external state, allowing the it to be in a
thermal state. For the generation of the states of Eq. (19), (28), and (32),
we do not require individual addressing of the ions.

In conclusion, we have proposed a scheme for generating entangled states for
two or more multi-level particles in both cavity QED and ion trap. In cavity
QED, our scheme does not require individual addressing of atoms in the
cavity. In cavity QED the scheme is insensitive to both cavity decay and
thermal field, which is of importance from the experimental point of view.
In ion trap, our scheme is insensitive to the thermal motion. Based on the
experiments reported in Refs. [10], [13], and [14], our scheme is realizable
with techniques presently available.

This work was supported by Fok Ying Tung Education Foundation 81008, the
National Fundamental Research Program Under Grant No. 2001CB309300, the
National Natural Science Foundation of China under Grant Nos. 60008003 and
10225421.

\end{document}